\documentclass{aa}
\usepackage[utf8]{inputenc}
\usepackage[varg]{txfonts}
\usepackage[breaklinks, colorlinks, citecolor=blue, linkcolor=blue]{hyperref}
\usepackage[usenames, dvipsnames]{color}
\usepackage{amsmath}
\usepackage{graphicx}
\usepackage[english]{babel}
\usepackage{siunitx}
\usepackage{float}
\usepackage{url}
\usepackage{longtable}
\usepackage{fancyhdr}
\usepackage{natbib}
\usepackage[normalem]{ulem}

\makeatletter
\def\instrefs#1{{\def\scsep{\def\scsep{,}}\@for\w:=#1\do{\scsep\ref{inst:\w}}}}
\renewcommand{\inst}[1]{\unskip$^{\instrefs{#1}}$}

\renewcommand*\aa@pageof{, page \thepage{} of \pageref*{LastPage}} 

\makeatother

\title{Obliquity measurement and atmospheric characterization of the WASP-74 planetary system}

\author{R.~Luque\inst{iac,ull} 
        \and
        N.~Casasayas-Barris\inst{iac,ull}
        \and
        H.~Parviainen\inst{iac,ull}
        \and
        G.~Chen\inst{pmo}
        \and
        E.~Pallé\inst{iac,ull}
        \and
        J.~Livingston\inst{utokyo}
        \and
        V.\,J.\,S.~Béjar\inst{iac,ull}
        \and
        N.~Crouzet\inst{esa}
        \and
        E.~Esparza-Borges\inst{ull}
        \and
        A.~Fukui\inst{deps-utokyo,iac}
        \and
        D.~Hidalgo\inst{iac,ull}
        \and
        Y.~Kawashima\inst{sron}
        \and
        K.~Kawauchi\inst{deps-utokyo}
        \and
        P.~Klagyivik\inst{dlr}
        \and
        S.~Kurita\inst{deps-utokyo}
        \and
        N.~Kusakabe\inst{nins,naoj}
        \and
        J.\,P.~de~Leon\inst{utokyo}
        \and
        A.~Madrigal-Aguado\inst{iac,ull}
        \and
        P.~Montañés-Rodríguez\inst{iac,ull}
        \and
        M.~Mori\inst{utokyo}
        \and
        F.~Murgas\inst{iac,ull}
        \and
        N.~Narita\inst{komaba,jst,nins,iac}
        \and
        T.~Nishiumi\inst{iguas,naoj}
        \and
        G.~Nowak\inst{iac,ull}
        \and
        M.~Oshagh\inst{iac,ull}
        \and
        M.~S\'anchez-Benavente\inst{iac,ull}
        \and
        M.~Stangret\inst{iac,ull}
        \and
        M.~Tamura\inst{utokyo,nins,naoj}
        \and
        Y.~Terada\inst{utokyo}
        \and
        N.~Watanabe\inst{iguas,nins}
        }

\institute{
\label{inst:iac}Instituto de Astrof\'isica de Canarias (IAC), 38205 La Laguna, Tenerife, Spain; \email{rluque@iac.es}
\and 
\label{inst:ull}Departamento de Astrof\'isica, Universidad de La Laguna (ULL), 38206, La Laguna, Tenerife, Spain
\and
\label{inst:pmo}Key Laboratory of Planetary Sciences, Purple Mountain Observatory, Chinese Academy of Sciences, Nanjing 210023, China
\and
\label{inst:utokyo}Department of Astronomy, The University of Tokyo, 7-3-1, Hongo, Bunkyo-ku, Tokyo, 113-0033, Japan
\and
\label{inst:esa}Science Support Office, Directorate of Science, European Space Research and Technology Centre (ESA/ESTEC), Keplerlaan 1, 2201 AZ Noordwijk, The Netherlands
\and
\label{inst:deps-utokyo}Department of Earth and Planetary Science, Graduate School of Science, The University of Tokyo, 7-3-1 Hongo, Bunkyo-ku, Tokyo 113-0033, Japan
\and
\label{inst:sron}SRON Netherlands Institute for Space Research, Sorbonnelaan 2, 3584 CA Utrecht, The Netherlands
\and
\label{inst:dlr}Institute of Planetary Research, German Aerospace Center, Rutherfordstrasse 2, 12489 Berlin, Germany
\and
\label{inst:nins}Astrobiology Center of NINS, 2-21-1, Osawa, Mitaka, Tokyo 181-8588, Japan
\and
\label{inst:naoj}National Astronomical Observatory of Japan, 2-21-1 Osawa, Mitaka, Tokyo 181-8588, Japan
\and
\label{inst:komaba}Komaba Institute for Science, The University of Tokyo, 3-8-1 Komaba, Meguro, Tokyo 153-8902, Japan
\and
\label{inst:jst}JST, PRESTO, 3-8-1 Komaba, Meguro, Tokyo 153-8902, Japan
\and
\label{inst:iguas}Department of Astronomical Science, The Graduated University for Advanced Studies, SOKENDAI, 2-21-1, Osawa, Mitaka, Tokyo, 181-8588 Japan
           }
           
\date{}

\abstract{

We present new transit observations of the hot Jupiter WASP-74~b ($T_\mathrm{eq} \sim$ 1860\,K) using the high-resolution spectrograph HARPS-N and the multi-colour simultaneous imager MuSCAT2. We refine the orbital properties of the planet and its host star, and measure its obliquity for the first time. The measured sky-projected angle between the stellar spin-axis and the planet's orbital axis is compatible with an orbit well-aligned with the equator of the host star ($\lambda = 0.77\pm0.99\,\mathrm{deg}$). We are not able to detect any absorption feature of H$\alpha$, or any other atomic spectral features, in its high-resolution transmission spectra due to low S/N at the line cores. Despite previous claims regarding the presence of strong optical absorbers such TiO and VO gases in the atmosphere of WASP-74~b, the new ground-based photometry combined with a reanalysis of previously reported observations from the literature shows a slope in the low-resolution transmission spectrum steeper than expected from Rayleigh scattering alone.

}


\begin{document}

\maketitle

\section{Introduction}

Metal oxides, such as TiO and VO, have been proposed to exist in the atmospheres of highly irradiated hot Jupiters, introducing thermal inversions in the temperature structure \citep{2003ApJ...594.1011H,2008ApJ...678.1419F}. However, these early theoretical predictions have barely been confidently confirmed by observations. The overall lack of TiO/VO detections in optical transmission spectroscopy then triggered several alternative theoretical interpretations, e.g., TiO/VO condensation \citep{2009ApJ...699.1487S}, stellar activity \citep{2010ApJ...720.1569K}, or high C/O ratio \citep{2012ApJ...758...36M}. 

The progress first appeared in the emission spectroscopy of so-called ultra-hot Jupiters \citep[UHJs, defined as gas giants with dayside temperatures hotter than $\sim$2200\,K;][]{2018A&A...617A.110P}. Both low- and high-resolution emission spectra of the UHJ WASP-33~b indicate the presence of TiO in its dayside atmosphere \citep{2015ApJ...806..146H,2017AJ....154..221N}. Later, both transmission and emission spectroscopy of another UHJ, WASP-121~b, revealed that the VO molecule is present in its atmosphere \citep{2016ApJ...822L...4E} and is responsible for the observed thermal inversion \citep{2017Natur.548...58E}, but the original tentative inference of TiO by multi-band photometry \citep{2016ApJ...822L...4E} was later ruled out by low-resolution transmission spectroscopy \citep{2018AJ....156..283E}. The first significant detection of TiO in the optical transmission spectrum came from the UHJ WASP-19~b \citep{2017Natur.549..238S}, although it is not confirmed by another independent work, suggesting that stellar contamination could introduce false positive TiO signatures \citep{2019MNRAS.482.2065E}. The search of TiO/VO in the optical transmission spectroscopy remains unresolved. An alternative explanation is that Fe is responsible for the thermal inversion, a claim that is gaining support with the recent detection of Fe I in the emission spectra of two ultra hot Jupiters, KELT-9~b \citep{Pino2020} and WASP-189~b \citep{Yan2020arXiv200702716Y}. 

Here we present a study of the system WASP-74 \citep{Hellier2015AJ....150...18H} using multi-colour photometry and high-resolution spectroscopy observations with MuSCAT2 (Multicolour Simultaneous Camera for studying Atmospheres of Transiting exoplanets) and HARPS-N (High Accuracy Radial velocity Planet Searcher in North hemisphere), respectively. WASP-74~b is a hot Jupiter in a 2-day orbit around a F9 star with a magnitude $V=9.75\,\mathrm{mag}$. With an equilibrium temperature of around $1900\,{\rm K}$, this planet remains very close to the UHJs region. \citet{Tsiaras2018AJ....155..156T} and \citet{Mancini2019MNRAS.485.5168M} measured its transmission spectra using the Wide Field Camera 3 (WFC3) on board of \textit{Hubble} Space Telescope ({\it HST}) and ground-based multi-band photometry, respectively. Their findings, although tentative, indicate a water-depleted atmosphere with strong optical absorbers such as TiO and VO. Here we revise these finding with our new observations and a re-analysis of the previously published data. 

Moreover, the time series of high resolution data during a transit allows us to investigate the architecture of the planetary system. During a transit, the planet blocks a moving portion of the stellar disk, and the corresponding RV of that stellar region is masked from the integrated stellar RV. This generates a RV anomaly during the transit which is known as the Rossiter–McLaughlin \citep[RM;][]{Rossiter-24, McLaughlin-24} effect. The RM signal has been used to estimate the projected spin–orbit angle ($\lambda$), the angle between the normal vector of the orbital plane and the stellar rotational spin-axis. So far a wide diversity of spin-orbit angles have been measured, ranging from aligned \citep{Winn10} to highly misaligned systems \citep{Addison-18}, and even retrograde planets \citep[e.g.,][]{Hebrard-11}. An statistically large sample of spin-orbit angle is essential for examining theories on planet formation and evolution \citep[e.g.,]{Winn-05, Triaud-10, Albrecht-12a, Triaud-18}. For instance, \citet{Winn2010ApJ...718L.145W} found that hot Jupiters have larger spin-orbit angles if they are orbiting around hot stars ($T_\mathrm{eff} > 6100\,\mathrm{K}$) . Several explanation were proposed for this trend which were connecting it to the star–planet tides that can align the planetary orbits with the stellar equator, and tides should be stronger for cooler stars. Here find that the WASP-74 spin orbit measurement is in line with this trend. 

This paper is organised as follows. In Section~\ref{sec:observations} we describe the multi-colour photometry and high-resolution spectroscopy observations. In Section~\ref{sec:stellarparams} we present the determination of the stellar parameters of the host star. In Section~\ref{sec:rossitter} we describe the retrieval of the spin orbit alignment of the system via RM measurements. In Section~\ref{sec:analysis} we analyse the atmosphere of the planet combining the photometric and spectroscopic data and discuss the absent of TiO/VO and the presence of Rayleigh scattering. Finally, in Section~\ref{sec:discussion} we present the summary of our results and conclusions.

\section{Observations} \label{sec:observations}

\begin{table*}[t]
\centering
\caption{Observing log of the WASP-74~b transit observations.}
\begin{tabular}{cccccccccccc}
\hline\hline
Tel. & Instrument & Date of  &Start  & End &  Filter & $T_\mathrm{exp}$ & $N_\mathrm{obs}$ & Airmass  & S/N\tablefootmark{a} \\
       & & observation & UT & UT & & [s] &               & & \ion{Na}{i} order  \\
\hline
\\[-1em]
TNG & HARPS-N & 2018-07-17 & 22:42 & 05:08 & - & 600 & 38 &  1.60$\rightarrow$1.15$\rightarrow$1.92 & 32-62 \\
\\[-1em]
TCS & MuSCAT2 & 2018-07-17 & 23:26 & 04:30 & $g$ $r$ $i$ $z$ & 8,8,8,8 & - & 1.38$\rightarrow$1.15$\rightarrow$1.55 & - \\
\\[.1em]
TNG & HARPS-N & 2018-08-01 & 21:11 & 04:19 & - & 600 & 42 & 1.88$\rightarrow$1.15$\rightarrow$2.04 &55-88 \\
\\[.1em]
TCS & MuSCAT2 & 2018-08-16 & 22:42 & 02:30 & $g$ $r$ $i$ $z$ & 20,20,20,20 & - & 1.30$\rightarrow$1.15$\rightarrow$1.55 & - \\
\\[.1em]
TNG & HARPS-N & 2018-08-31 & 20:19 & 02:14 & - & 600 & 35 & 1.41$\rightarrow$1.15$\rightarrow$1.96 & 46-64 \\
\\[-1em]
TCS & MuSCAT2 & 2018-08-31 & 20:49 & 00:55 & $g$ $r$ $i$ $z$ & 10,8,10,15 & - & 1.32$\rightarrow$1.15$\rightarrow$1.40 & - \\
\\[.1em]
TCS & MuSCAT2 & 2019-06-24\tablefootmark{b} & 00:21 & 03:24 & $g$ $r$ $i$ & 12,8,18 & - & 1.55$\rightarrow$1.15$\rightarrow$1.16 & - \\
\\[-1em]
\\[-1em]
\hline
\end{tabular}\\
\tablefoot{
\tablefoottext{a}{Averaged S/N per extracted pixel calculated in the \ion{Na}{i} order ($590~{\rm nm}$) for each night.} \tablefoottext{b}{Partial transit, discarded for the joint photometric analysis.}
}
\label{Tab:Obs}
\end{table*}

\subsection{Multi-colour photometry}

We observed four transits (three full, one partial) of WASP-74~b with the MuSCAT2 multi-colour imager \citep{MuSCAT2} installed in Telescopio Carlos S\'anchez (TCS) located at the Teide Observatory in Tenerife, Spain. Observations were carried out simultaneously in four colours ($g$, $r$, $i$, $z$) in the three full transits (2018-07-17, 2018-08-16, and 2018-08-31) and only in three colours ($g$, $r$, $i$) for the partial one (2019-06-24) with a pixel scale of 0.44\arcsec\,pix$^{-1}$. The $z$ band observations are missing in this transit due to a problem with its CCD, which was under maintenance. A summary of the key properties for each of the nights is presented in Table~\ref{Tab:Obs}.

The reduction of the multi-colour photometry data was performed with a dedicated MuSCAT2 pipeline including bias and flat-field correction. In a nutshell, it calculates aperture photometry for a set of comparison stars and photometry aperture sizes, and creates the final relative light curves via global optimisation of the posterior density for a model consisting of a transit model (with quadratic limb darkening coefficients), apertures, comparison stars, and a linear baseline model with the airmass, seeing, x- and y-centroid shifts, and the sky level as covariates (see \citealt{Parviainen2019A&A...630A..89P} for details). 

\citet{Mancini2019MNRAS.485.5168M} collected broad-band photometry in several filters of WASP-74~b in order to determine the observational transmission spectrum of the planet. Their dataset comprises a total of 18 light curves from 11 different transits between 2015 and 2017 in the following passbands: Bessell $U$, Johnson $B$, Sloan $g'$, $r'$, $i'$, and $z'$, Bessell $I$, Cousins $I$, and near-infrared $J$, $H$, and $K$ bands. Observations were carried out in the following telescopes: Calar Alto 1.23-m telescope (one transit in Johnson $B$ and another in Cousins $I$), Danish 1.54-m telescope (seven transits in Bessell $I$ and two more in Bessell $U$), and the GROND (Gamma-Ray Burst Optical/Near-Infrared Detector) multi-colour imager at the MPG 2.2-m telescope in La Silla, Chile (one transit in $g'$, $r'$, $i'$, $z'$, $J$, $H$, and $K$). Besides, WASP-74~b was observed with the HST/WFC3 camera by \citet{Tsiaras2018AJ....155..156T} for measuring the transmission spectra of a sample of hot Jupiters from 1.1 to 1.7\,$\mu$m and with \textit{Spitzer}/IRAC in 3.6 and 4.5\,$\mu$m (PI: Deming) for a statistical study of secondary eclipses of hot Jupiters by \citet{Garhart2020AJ....159..137G}.

While we use the \textit{HST} observations as presented in \citet{Tsiaras2018AJ....155..156T}, we perform our own photometric analysis of the \textit{Spitzer} observations. As in \citet{Livingston2019AJ....157..102L}, we extract the \textit{Spitzer}  light curves following the approach taken by \citet{Knutson2012ApJ...754...22K,Beichman2016ApJ...822...39B} and select the circular aperture size that minimises the combined uncorrelated and correlated noise (2.2\,pix), as measured by the standard deviation and $\beta$ factor \citep{Pont2006MNRAS.373..231P,Winn2008ApJ...683.1076W}. Then, we model jointly the transit and systematics inherent to the \textit{Spitzer}  light curves using the pixel-level decorrelation method \citep{Deming15}, which uses a linear combination of (normalised) pixel light curves to model the effect of point-spread function (PSF) motion on the detector coupled with intrapixel gain variations. The \textit{Spitzer} PLD-corrected light curves are shown in Fig.~\ref{fig:transitLC_all} and fitted jointly with the remaining photometry in Sect.~\ref{subsec:lightcurve}.

\subsection{High-resolution spectroscopy}

Three transits of WASP-74~b were observed using the HARPS-N spectrograph \citep{HARPSN12003Msngr.114...20M,HARPSN22012SPIE.8446E..1VC}, mounted on the $3.58$-m Telescopio Nazionale Galileo (TNG), located at the Observatorio del Roque de los Muchachos in La Palma, Spain. Two of them were simultaneously observed with MuSCAT2. HARPS-N covers the optical wavelength regime between 0.38\,$\mu$m and 0.69\,$\mu$m with a spectral resolution of $\mathcal{R}=115\,000$. The observations were performed exposing continuously before, during, and after the transit, using an exposure time of $600\,{\rm s}$. The signal-to-noise (S/N), calculated as an average of the S/N per pixel in the \ion{Na}{i} order ($590~{\rm nm}$), ranges from 55--88 in the second night and 32--64 in the first and third nights. In all three cases, we used fiber B to monitor possible sky emission during the night. Details on the observations are presented in Table~\ref{Tab:Obs}.

HARPS-N observations were reduced using the HARPS-N Data reduction Software (DRS), version $3.7$ \citep{HARPSNDRS12014SPIE.9147E..8CC,HARPSNDRS22014ASPC..485..435S}. After computing the wavelength calibration solution, the DRS combines and resamples the two-dimensional echelle spectra with wavelength step of $0.01\,\mathrm{\AA}$ into a one-dimensional spectrum. The final spectra are referred to the barycentric rest frame and standard air wavelengths are used.

\section{Stellar parameters} \label{sec:stellarparams}

\begin{table}
\centering
{\renewcommand{\arraystretch}{1.0}
 \footnotesize
\caption{Stellar parameters of WASP-74.} \label{tab:star}
\begin{tabular}{lcr}
\hline\hline
\noalign{\smallskip}
Parameter                               & Value                 & Reference \\ 
\noalign{\smallskip}
\hline
\noalign{\smallskip}
Name                        & WASP-74                   & \citet{Hellier2015AJ....150...18H}     \\
\noalign{\smallskip}
\multicolumn{3}{c}{Coordinates and spectral type}\\
\noalign{\smallskip}
$\alpha$                    & 20:18:09.32               & {\it Gaia} DR2     \\
$\delta$                    & $-$01:04:32.6             & {\it Gaia} DR2     \\
Spectral type               & F\,9                      & \citet{Hellier2015AJ....150...18H}          \\
\noalign{\smallskip}
\multicolumn{3}{c}{Magnitudes}\\
\noalign{\smallskip}
$B$ [mag]                   & $10.39\pm0.04$        & Tycho-2       \\
$V$ [mag]                   & $~9.75\pm0.03$        & Tycho-2       \\
$G$ [mag]                   & $9.5723\pm0.0003$     & {\it Gaia} DR2       \\
$J$ [mag]                   & $8.548\pm0.037$       & 2MASS       \\
$H$ [mag]                   & $8.286\pm0.018$       & 2MASS       \\
$K_s$ [mag]                 & $8.221\pm0.023$       & 2MASS       \\
\noalign{\smallskip}
\multicolumn{3}{c}{Parallax and kinematics}\\
\noalign{\smallskip}
$\pi$ [mas]                     & $6.673\pm0.051$       & {\it Gaia} DR2             \\
$d$ [pc]                        & $149.8\pm1.1$         & {\it Gaia} DR2             \\
$\mu_{\alpha}\cos\delta$ [$\mathrm{mas\,yr^{-1}}$]      & $1.350 \pm 0.082$     & {\it Gaia} DR2         \\
$\mu_{\delta}$ [$\mathrm{mas\,yr^{-1}}$]                & $-64.60 \pm 0.06$     & {\it Gaia} DR2     \\
$V_r [\mathrm{km\,s^{-1}}]$     & $-15.32 \pm 0.27$     & {\it Gaia} DR2     \\
                                & $-15.45 \pm 0.35$     & This work     \\
\noalign{\smallskip}
\multicolumn{3}{c}{Photospheric parameters}\\
\noalign{\smallskip}
$T_{\mathrm{eff}}$ [K]      & $5883 \pm 57$             & This work         \\
                            & $5984 \pm 57$             & \citet{Mancini2019MNRAS.485.5168M}  \\
                            & $5990 \pm 110$            & \citet{Hellier2015AJ....150...18H}  \\
$\log g_{\star}$            & $4.05 \pm 0.02$           & This work         \\
                            & $4.141\pm0.011\pm0.004$   & \citet{Mancini2019MNRAS.485.5168M}         \\
                            & $4.180\pm0.018$           & \citet{Hellier2015AJ....150...18H}         \\
{[Fe/H]}                    & $0.38\pm0.03$             & This work         \\
                            & $0.34\pm0.02$             & \citet{Mancini2019MNRAS.485.5168M}         \\
                            & $0.39\pm0.13$             & \citet{Hellier2015AJ....150...18H}         \\
$v \sin I_{\star}$ [$\mathrm{km\,s^{-1}}$]      & $6.13 \pm 0.21$     & This work         \\
                            & $6.03\pm0.19$             & \citet{Mancini2019MNRAS.485.5168M}         \\
                            & $4.1\pm0.8$               & \citet{Hellier2015AJ....150...18H}         \\
\noalign{\smallskip}
\multicolumn{3}{c}{Derived physical parameters}\\
\noalign{\smallskip}
$M_{\star}$ [M$_{\odot}$]   & $1.236 \pm 0.026$         & This work         \\
                            & $1.191\pm0.023\pm0.030$   & \citet{Mancini2019MNRAS.485.5168M}         \\
                            & $1.48\pm0.12$             & \citet{Hellier2015AJ....150...18H}         \\
$R_{\star}$ [R$_{\odot}$]   & $1.444 \pm 0.044$         & This work         \\
                            & $1.536\pm0.022\pm0.013$   & \citet{Mancini2019MNRAS.485.5168M}         \\
                            & $1.64\pm0.05$             & \citet{Hellier2015AJ....150...18H}         \\
Age [Gyr]                   & $3.49 \pm 0.65$           & This work         \\
                            & $4.2^{+0.4+1.6}_{-0.4-2.0}$   & \citet{Mancini2019MNRAS.485.5168M}         \\
                            & $2.0^{+1.6}_{-1.0}$       & \citet{Hellier2015AJ....150...18H}         \\
\noalign{\smallskip}
\hline
\end{tabular}}
\tablebib{
    {\it Gaia} DR2: \citet{GaiaDR2}; Tycho-2: \citet{Tycho-2}; 2MASS: \citet{2MASS}.
}
\end{table}

We used the Zonal Atmospheric Stellar Parameters Estimator \citep[ZASPE;][]{ZASPE} code to determine the atmospheric stellar parameters of WASP-74. The parameters were obtained with a high S/N spectrum built by co-adding all HARPS-N out-of-transit observations. In summary, ZASPE matches the observed stellar spectrum via least-squares minimisation against a grid of synthetic spectra in the spectral regions most sensitive to changes in $T_{\rm eff}$, $\log g_\star$, and [Fe/H]. Then, to derive the physical parameters of the star, we used \texttt{PARAM~1.3}\footnote{\url{http://stev.oapd.inaf.it/cgi-bin/param_1.3}.}, a web interface for Bayesian estimation of stellar parameters using the \texttt{PARSEC} isochrones from \citet{PARSEC}. The required input is the effective temperature and metallicity of the star, determined spectroscopically, together with its apparent visual magnitude and parallax.

We derive an effective temperature of $T_{\rm eff}=5883\pm57\,\mathrm{K}$, a stellar mass of $M_\star = 1.236 \pm 0.026\,\mathrm{M_{\odot}}$, and a radius of $R_\star = 1.444 \pm 0.044\,\mathrm{R_{\odot}}$, in fairly good agreement with the most up-to-date values reported in \citet{Mancini2019MNRAS.485.5168M}. The stellar models constrain the age of the star to be $3.49 \pm 0.65\,\mathrm{Gyr}$. We stress that the uncertainties on the derived parameters are internal to the stellar models used and do not include systematic uncertainties related to input physics. All derived values and previous ones reported in the literature can be found in Table~\ref{tab:star}.

\section{Planetary obliquity} 
\label{sec:rossitter}

\begin{figure*}
\centering
\includegraphics[width=\hsize]{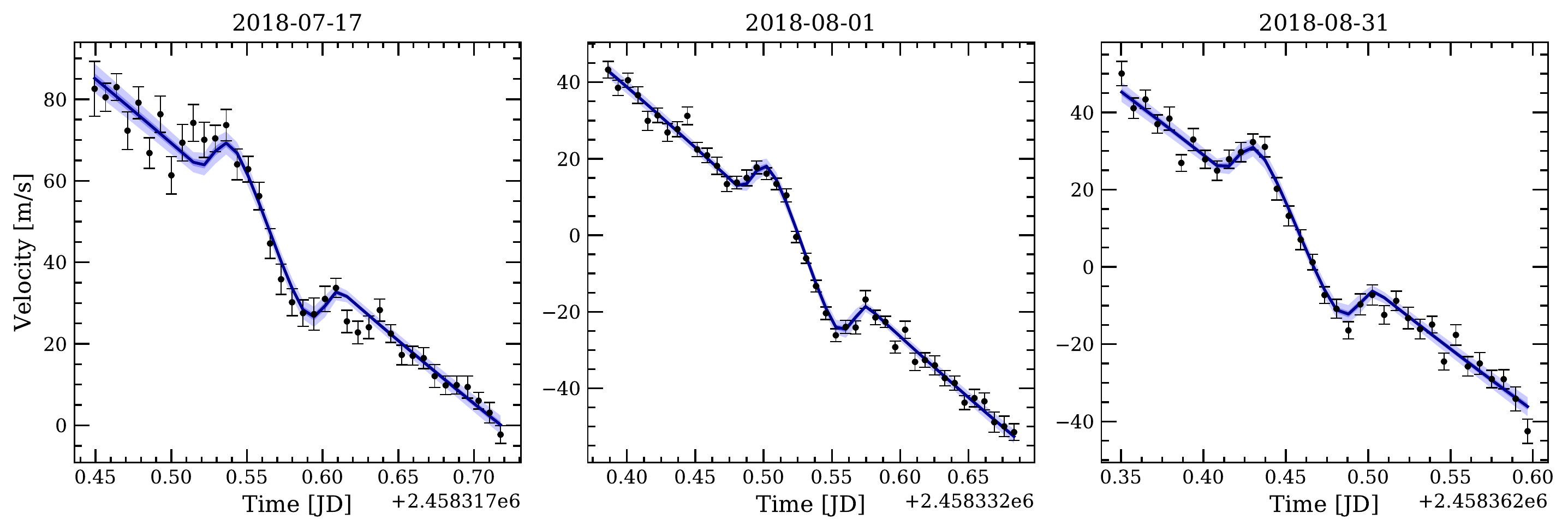}\\
\caption{Stellar radial velocities of WASP-74 for each individual night. The radial velocity measurements are shown in black dots. In cyan we show the best fit model obtained with the MCMC procedure.} \label{fig:RM_indiv}
\end{figure*}

\begin{figure}
\centering
\includegraphics[width=0.9\hsize]{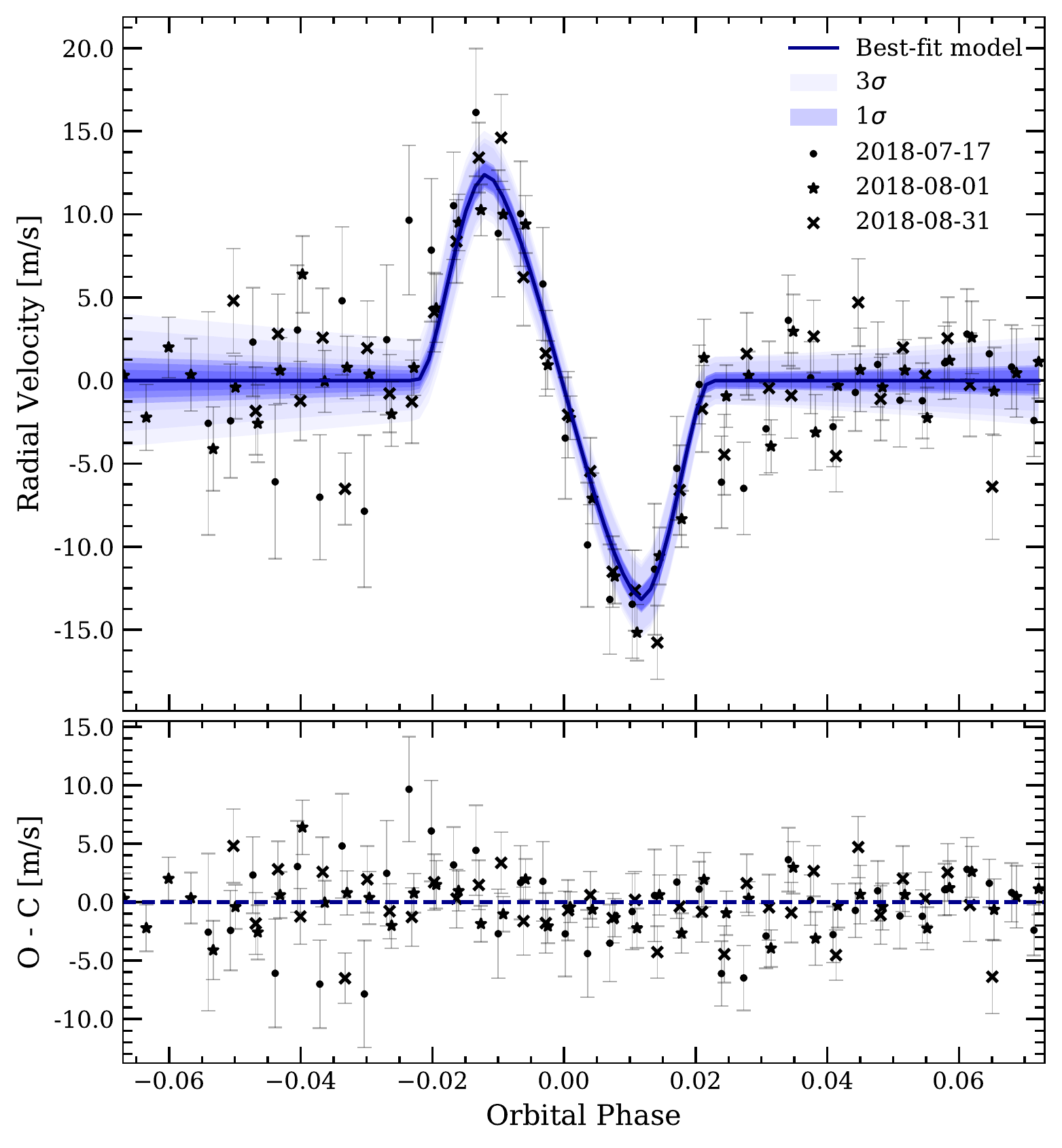}
\caption{RM effect during the transit of WASP-74~b after being detrended (top panel), and residuals between the data and model (bottom panel). In different symbols we show the radial velocities from the three different data sets. The dark blue line corresponds to the best-fit RM model. In light blue we show the $1\sigma$ and $3\sigma$ uncertainties of the model.} \label{fig:RM_all}
\end{figure}

\begin{table}
\centering
{\renewcommand{\arraystretch}{1.2}
\footnotesize
\caption{Estimates for the system parameters derived from RM effect analysis.}
\begin{tabular}{llr}
\hline\hline
\noalign{\smallskip}
Parameter & Unit & Value \\
\noalign{\smallskip}
\hline
\noalign{\smallskip}
$\lambda$ & [deg] & $0.77\pm0.99$\\ 
$\Omega$ & [rad d$^{-1}$] & $0.503^{+0.041}_{-0.046}$\\ 
$v\sin I_{\star}$ & [km s$^{-1}$] & $5.85\pm0.50$\\ 
$T_0$ & [JD] &  $2457173.86703\pm0.00075$\\ 
$\epsilon$ & ... & $0.88^{+0.07}_{-0.11}$\\ 
$K_{\star,1}$ & [m s$^{-1}$] & $110.8\pm2.3$\\ 
$K_{\star,2}$ & [m s$^{-1}$] & $113.0\pm1.3$\\ 
$K_{\star,3}$ & [m s$^{-1}$] & $115.0\pm2.1$\\ 
$\gamma_1$\tablefootmark{a} & [m s$^{-1}$] & $48.21\pm0.62$\\ 
$\gamma_2$ & [m s$^{-1}$] & $-3.11\pm0.36$\\ 
$\gamma_3$ & [m s$^{-1}$] & $9.57\pm0.47$\\ 
\noalign{\smallskip}
\hline
\end{tabular}
\tablefoot{\tablefoottext{a}{The super scripts $1$, $2$, and $3$ refer to the results obtained for nights 2018-07-17, 2018-08-01 and 2018-08-31, respectively.}}
\label{tab:RM_res}}
\end{table}

\begin{figure}
\centering
\includegraphics[width=0.98\hsize]{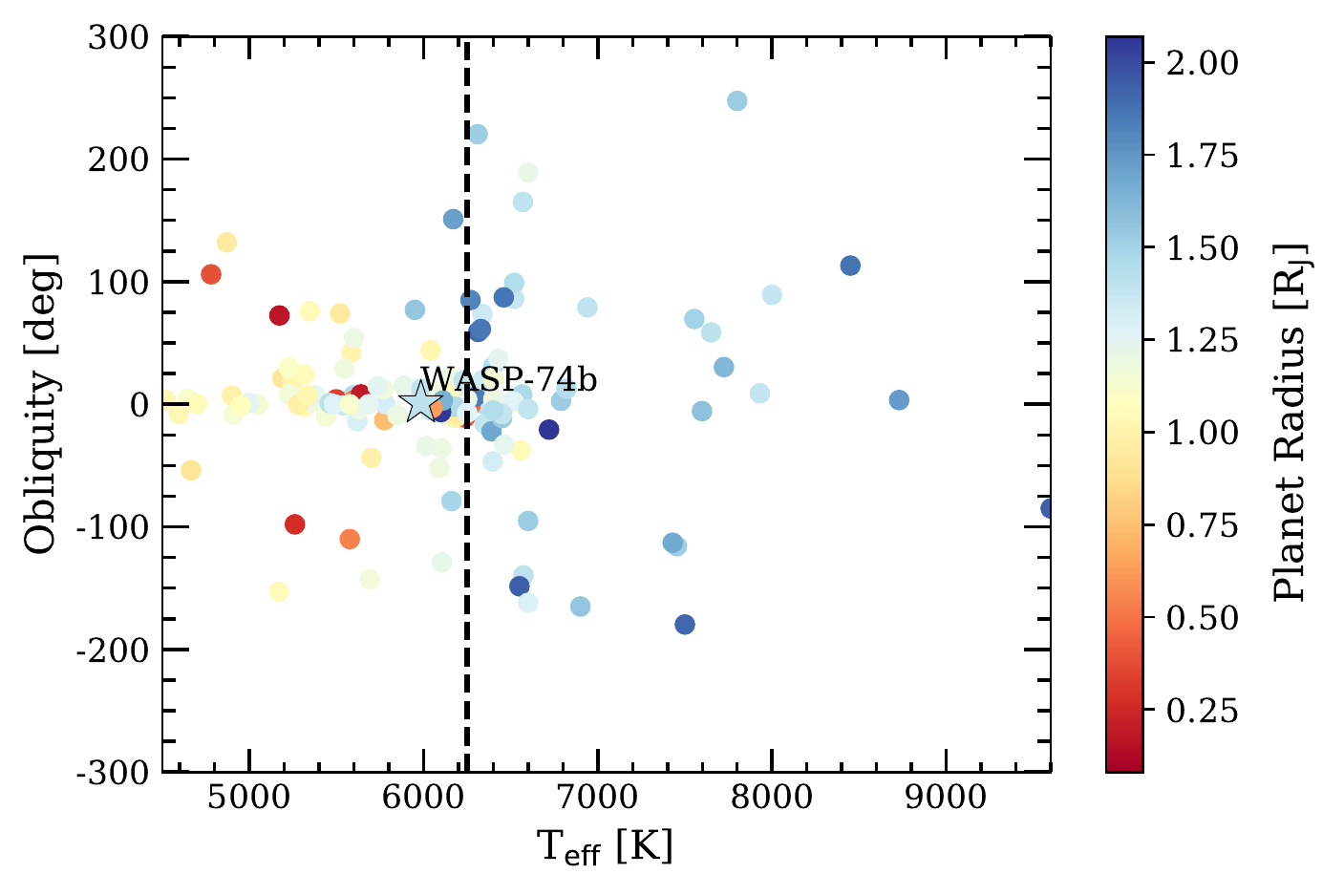}
\caption{Measurements of orbital obliquity for the known transiting extrasolar planetary systems and brown dwarf companions in front of the effective temperature of the host star (dots), extracted from TEPCat orbital obliquity catalogue \citep{TEPCat}. In the colour bar we present the planetary radius. The star symbol corresponds to WASP-74~b's spin-orbit measurement. The black-dashed vertical line marks the $6250~{\rm K}$ effective temperature transition from \citet{Winn2010}.} \label{fig:obl_context}
\end{figure}

The radial velocities of the three nights were computed using \texttt{serval} \citep{SERVAL}, which uses least-squares fitting with a high S/N template to compute the radial velocities. The template is created by co-adding all the out-of-transit spectra of the star. The Rossiter-McLaughlin (RM) effect is clearly observed in the extracted radial velocities of each individual night (Fig.~\ref{fig:RM_indiv}).

In order to measure the obliquity ($\lambda$) of the system, we fit a RM model to the radial velocity (RV) data by using the the Markov chain Monte Carlo (MCMC) algorithm implemented in {\tt emcee} \citep{emcee2013PASP..125..306F}. We use two different RV contributions to build our model: the RM effect and a circular orbit. Both models are implemented in {\tt PyAstronomy} \citep{PyAstronomy2019ascl.soft06010C} as {\tt modelSuite.RmcL} and {\tt modelSuite.radVel}, respectively. The model containing the RM effect depends on the orbital period ($P$), the transit epoch ($T_0$), the planet-to-star radius ratio ($R_p/R_{\star} \equiv k$), the angular rotation velocity of the host star ($\Omega$), the linear limb-darkening coefficient ($\epsilon$), the inclination of the orbit ($i$), the inclination of the stellar rotation axis ($I_{\star}$), the sky-projected angle between the stellar rotation axis and the normal of planetary orbit plane ($\lambda$) and the scaled semi-major axis ($a/R_{\star} \equiv a_\mathrm{s}$). On the other hand, the circular obit RV contribution depends on $P$, T$_c$, the stellar velocity semi-amplitude ($K_{\star}$), and the offset with respect to the null RV ($\gamma$). 

As presented in previous studies (e.g., \citealt{Casasayas2017}), in the fitting procedure, $I_{\star}$ to $90~{\rm deg}$, while  $T_0$, $a_\mathrm{s}$, $k$, and $R_{\star}$ are fixed to the values derived in Sections~\ref{sec:stellarparams} and \ref{subsec:lightcurve}. The other parameters ($\Omega$, $\epsilon$, $\lambda$, $K_{\star}$ and $\gamma$) remain free. The RV information from the three nights is jointly fitted, considering that $T_0$, $\Omega$, $\epsilon$, $\lambda$ are shared parameters. On the other hand, the offset between the model and the data ($\gamma$) can vary from night to night as, additionally to the system velocity, the RV information is given with possible instrumental and stellar activity effects. $K_{\star}$ could also be affected by activity and become different for different nights \citep{MahmoudActivity2018A&A...619A.150O}. For this reason, we fit one different $\gamma$ and $K_{\star}$ per night (called $\gamma_1$, $\gamma_2$, $\gamma_3$ and $K_{\star,1}$, $K_{\star,2}$, $K_{\star,3}$, respectively).

We analyse the system using 50 walkers and a total of $10^6$ steps and checked their convergence using the Gelman-Rubin statistic. Adequate convergence was considered when the Gelman–Rubin potential scale reduction factor dropped to within 1.03. Each step is initialised at a random point near the measured values from literature. $\lambda$ is constrained to $\pm180\,{\rm deg}$, $\epsilon$ between $0.5$ and $1.0$ (using the {\tt ldtk} \citep{LDTk} we estimate a linear limb-darkening coefficient of $0.71$ in the HARPS-N wavelength coverage), and $\Omega$ between $0.3$ and $0.9$\,rad\,d$^{-1}$, which is translated to $v\sin I_{\star}$ limited between $3.7$ and $11.1$\,km\,s$^{-1}$ (\citet{Mancini2019MNRAS.485.5168M} measured a $v\sin I_{\star}=6.03$\,km\,s$^{-1}$). The median values of the posteriors are adopted as the best-fit values, and their error bars correspond to the $1\sigma$ statistical errors at the corresponding percentiles. The individual RM curves can be observed in Fig.~\ref{fig:RM_indiv}. The detrended data of all nights with the best-fit model are presented in Fig.~\ref{fig:RM_all}.

With the joint fit of the three nights, we measure a spin-orbit angle of $0.8\pm1.0$\,deg, meaning an aligned system. The angular rotation velocity is measured to be of $0.50\pm0.04\,{\rm rad\,d^{-1}}$ and $v\sin I_{\star} = 5.85\pm0.50$\,km\,s$^{-1}$, consistent with our spectroscopically derived results from Table~\ref{tab:star}, and with the results obtained by \citet{Mancini2019MNRAS.485.5168M}. The $v\sin I_{\star}$ value derived from the RM fitting differs less than $2\sigma$ from \citet{Hellier2015AJ....150...18H} results. Also, the $K_{\star}$ values of the individual nights are consistent among themselves and with the value reported in \citet{Hellier2015AJ....150...18H}, pointing to a low level of stellar activity \citep{MahmoudActivity2018A&A...619A.150O}. This is supported by the absence of spot-crossing events in the MuSCAT2 simultaneous multi-colour photometric observations during HARPS-N first and third transits. Although we cannot assess the impact associated to un-occulted spots affecting both RM and transit observations, we can be confident that our $\lambda$ and $v\sin I_{\star}$ determinations are not significantly misestimated. There are two reasons which support this claim; first the results obtained with the joint fit are consistent with the results obtained when fitting each night independently. Second combining three RMs, as was demonstrated in \citet{MahmoudActivity2018A&A...619A.150O}, are sufficient to mitigate and minimise the influence of stellar activity on estimated $\lambda$ and $v\sin I_{\star}$. However, as presented in \citet{Cegla2016} and \citet{Bourrier2017}, the spin-orbit and $v\sin I_{\star}$ measurements performed using the classical RM could be significantly biased due to variations in the shape of the local cross-correlation functions (CCFs). 

In Figure~\ref{fig:obl_context} we show the obliquity measurements for known transiting planets (from TEPCat orbital obliquity catalogue; \citealt{TEPCat}) with respect to the effective temperature of their host stars. As presented in \citet{Winn2010}, we confirm that most of the planets orbiting stars with effective temperatures lower than ${\sim 6200}~{\rm K}$ are in aligned systems, while those planets orbiting hotter stars tend to form misaligned systems with a higher frequency. In this context, WASP-74's system is in agreement with this trend, located at the low-obliquity region. As explained in this same study, this could be the result of the interaction between the planetary orbit inside the convective zone of cool stars, which due to tidal dissipation realign the star-planet system. 

\section{Atmospheric characterization} 
\label{sec:analysis}

\subsection{Multi-colour light curve analysis} \label{subsec:lightcurve}

\begin{figure*}
\centering
\includegraphics[width=\textwidth]{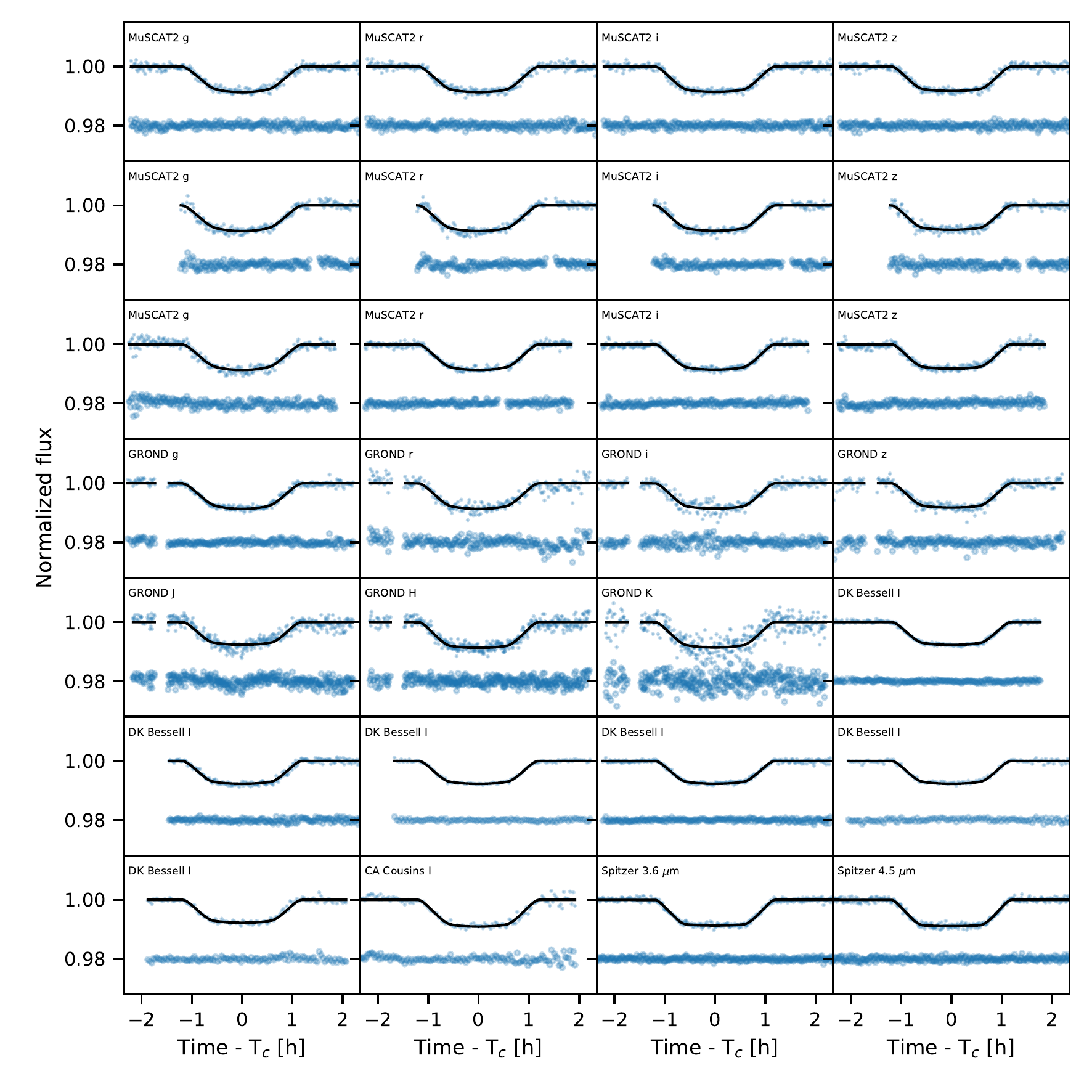}
\caption{Light curves of three full transits of WASP-74~b observed with MuSCAT2, a transit observed with GROND,  seven transits observed with DK, two transits observed with CA, and two transits observed with \textit{Spitzer}. A transit model in black corresponding to the median of the model parameter posteriors. 
} \label{fig:transitLC_all}
\end{figure*}

We model the three full transits from MuSCAT2 jointly with the two  \textit{Spitzer}/IRAC (3.6\,$\mu$m and 4.5\,$\mu$m channels), seven GROND (Sloan $g'$, $r'$, $i'$, $z'$, $J$, $H$, and $K$ passbands), seven Danish 1.54-m Telescope (Bessell $I$ passband), and one Calar Alto 1.23-m (Cousins $I$ passband) light curves presented in \citep{Mancini2019MNRAS.485.5168M}\footnote{Kindly provided to us by L.~Mancini, personal communication.}. We exclude the two Danish Bessell $U$ light curves due to their short pre- and post-transit baselines and strong correlated noise that cannot be sufficiently accounted for using the data available, and the CA Johnson $B$ light curve because of partial transit coverage (we include partial transits only when we have light curves with full transit coverage in the same passband).

We carry out the light curve analysis of the data in a Bayesian framework following \citet{Parviainen2018haex.bookE.149P}. First, we construct a flux model to reproduce both the transit and the light curve systematics. Then, we define a noise model to incorporate possible stochastic variability in the observations and combine it with the flux model and the observations to define the likelihood. Using MCMC sampling, we estimate the joint parameter posterior distributions after defining the priors on the model parameters. The analyses were carried out with a custom Python code based on \texttt{PyTransit} \citep{PyTransit}, \texttt{LDTk} \citep{LDTk}, \texttt{emcee} \citep{emcee2013PASP..125..306F}, and other standard Python libraries for astrophysics and scientific computing.

The four parameters describing the orbital geometry of the planet (zero epoch $T_0$, orbital period $P$, stellar density $\rho_\star$, and impact parameter $b$) are independent of passband or per-light-curve systematics, and thus all the light curves constrain the posterior distributions of these parameters. The radius ratio $k$ and the limb darkening coefficients are wavelength dependent, and thus all the light curves observed in a given passband constrain the posterior densities of these parameters in that passband. Finally, systematics are modelled using a linear combination of state vectors where the number of covariates varies from dataset to dataset (in most cases at least airmass and x- and y- centroid shifts are available). Each covariate is associated with a free coefficient, and the coefficient posteriors are constrained only by the information in the light curve modelled by the covariate.

We have eleven separate passbands ($g$, $r$, $i$, Cousins $I$, Johnson $I$, $z$, $J$, $H$, $K$, 3.6\,$\mu$m and 4.5\,$\mu$m), so we end up with eleven radius ratio parameters and 22 parameters for the quadratic stellar limb darkening model. The area ratios have wide uniform priors that do not constrain the posterior, but the limb darkening coefficients have loosely constraining normal priors created using \texttt{LDTk} (the LDTk-derived prior standard deviation is multiplied by 10) for all the passbands except $J$, $H$, and $K$. We set uniform priors for those passband limb darkening coefficients that constrain the coefficients to smaller values than the LDTk-derived prior for the $z$ band.

The linear baseline model contributes 105 free parameters to the model in total. The MuSCAT2 light curves have four covariates each (airmass, x- and y- shifts, and aperture entropy that works as a proxy for FWHM) and the GROND light curves have five covariates each (x- and y-shifts, x- and y-FWHMs, and airmass). The publicly available Danish 1.54-m and Calar Alto 1.23-m light curves do not include covariate information.

We first analyse the MuSCAT2 data and GROND data separately to test whether the two datasets agree with each other, and then run the analysis using the MuSCAT2, GROND, Danish, Calar Alto, and \textit{Spitzer} light curves jointly (the DK and CA light curves in $I$ band cannot be directly compared with MuSCAT2 and GROND $i$ filter due to broader wavelength coverage). The MuSCAT2 analysis agrees well with the GROND analysis, although the radius ratio posteriors of the GROND data have larger uncertainties that can be attributed to a passband-dependent instrumental effect (discontinuity) very close to the transit centre that cannot be sufficiently modelled by the linear baseline model.

We adopt the results from the full joint analysis as our final result and present them in Table~\ref{tab:parameters}, and we also plot all the light curves in Fig.~\ref{fig:transitLC_all}.

\begin{table}
\centering
{\renewcommand{\arraystretch}{1.2}
 \normalsize
\caption{Stellar and planetary parameters derived from the multi-colour joint transit analysis of WASP-74~b.}
\label{tab:parameters} 
\begin{tabular}{llr}        
\hline\hline
\noalign{\smallskip}
Parameter & Unit & Value\tablefootmark{(a)} \\
\noalign{\smallskip}
\hline
\noalign{\smallskip}
\multicolumn{3}{c}{Ephemeris} \\
\noalign{\smallskip}
$T_0$       & [BJD]     & $ 2457173.871756 \pm 9 \times 10^{-5}$     \\
$P$         & [d]       & $ 2.13775138 \pm 2.4\times 10^{-7} $        \\
$T_{14}$    & [h]       & $ 2.38 \pm 0.02 $ \\
\noalign{\smallskip}
\multicolumn{3}{c}{Fitted parameters} \\
\noalign{\smallskip}
$k_\mathrm{g}$ & & 0.09943 $\pm$ 0.00081 \\
$k_\mathrm{r}$ & & 0.09762 $\pm$ 0.00072 \\
$k_\mathrm{i}$ & & 0.09627 $\pm$ 0.00063 \\
$k_\mathrm{s}$ & & 0.09376 $\pm$ 0.00064 \\
$k_\mathrm{CI}$ & & 0.09844 $\pm$ 0.00119 \\
$k_\mathrm{BI}$ & & 0.09128 $\pm$ 0.00040 \\
$k_\mathrm{z}$ & & 0.09376 $\pm$ 0.00064 \\
$k_\mathrm{J}$ & & 0.08976 $\pm$ 0.00158 \\
$k_\mathrm{H}$ & & 0.09544 $\pm$ 0.00154 \\
$k_\mathrm{K}$ & & 0.09477 $\pm$ 0.00246 \\
$k_\mathrm{3.6\mu m}$ & & 0.09406 $\pm$ 0.00070 \\
$k_\mathrm{4.5\mu m}$ & & 0.09509 $\pm$ 0.00073 \\
$a_\mathrm{s}$          & [R$_\star$]   & $ 4.97 \pm 0.03 $ \\
$b$                     &               & $ 0.84 \pm 0.01 $ \\
$\rho_\star$            & $[\mathrm{g\,cm^{-3}}]$ & $ 0.509 \pm 0.01 $\\
\noalign{\smallskip}
\multicolumn{3}{c}{Derived parameters\tablefootmark{(b)}} \\
\noalign{\smallskip}
$R_\mathrm{p,g}$ & $[R_J]$ & 1.429 $\pm$ 0.045 \\
$R_\mathrm{p,r}$ & $[R_J]$ & 1.403 $\pm$ 0.044 \\
$R_\mathrm{p,i}$ & $[R_J]$ & 1.383 $\pm$ 0.043 \\
$R_\mathrm{p,z}$ & $[R_J]$ & 1.348 $\pm$ 0.042 \\
$R_\mathrm{p,CI}$ & $[R_J]$ & 1.415 $\pm$ 0.046 \\
$R_\mathrm{p,BI}$ & $[R_J]$ & 1.312 $\pm$ 0.040 \\
$R_\mathrm{p,J}$ & $[R_J]$ & 1.290 $\pm$ 0.045 \\
$R_\mathrm{p,H}$ & $[R_J]$ & 1.372 $\pm$ 0.047 \\
$R_\mathrm{p,K}$ & $[R_J]$ & 1.362 $\pm$ 0.055 \\
$R_\mathrm{p,3.6\mu m}$ & $[R_J]$ & 1.352 $\pm$ 0.042 \\
$R_\mathrm{p,4.5\mu m}$ & $[R_J]$ & 1.367 $\pm$ 0.043 \\
$a$                     & [AU]                  & $ 0.0334 \pm 0.001 $\\
$i$                     &[deg]                  & $80.32 \pm 0.09$ \\
$T_{\mathrm{eq}}$       & [K]                   & $ 1865 \pm 20 $ \\
\noalign{\smallskip}
\hline       
\noalign{\smallskip}
\end{tabular}
\tablefoot{ 
\tablefoottext{a}{The estimates correspond to the posterior median ($P_{50}$) with $1 \sigma$ uncertainty estimate based on the 16th and 84th posterior percentiles ($P_{16}$ and $P_{84}$, respectively) for symmetric, approximately normal posteriors. For asymmetric, unimodal, posteriors, the estimates are $P_{50}{}^{P_{84}-P_{50}}_{P_{16}-P_{50}}$.}
 \tablefoottext{b}{The derived planetary parameters are based on the stellar parameters shown in Table~\ref{tab:star}}.
}
}
\end{table}

\subsection{Low-resolution transmission spectrophotometry} \label{subsec:broadband}

\begin{figure*}
\centering
\includegraphics[width=0.98\hsize]{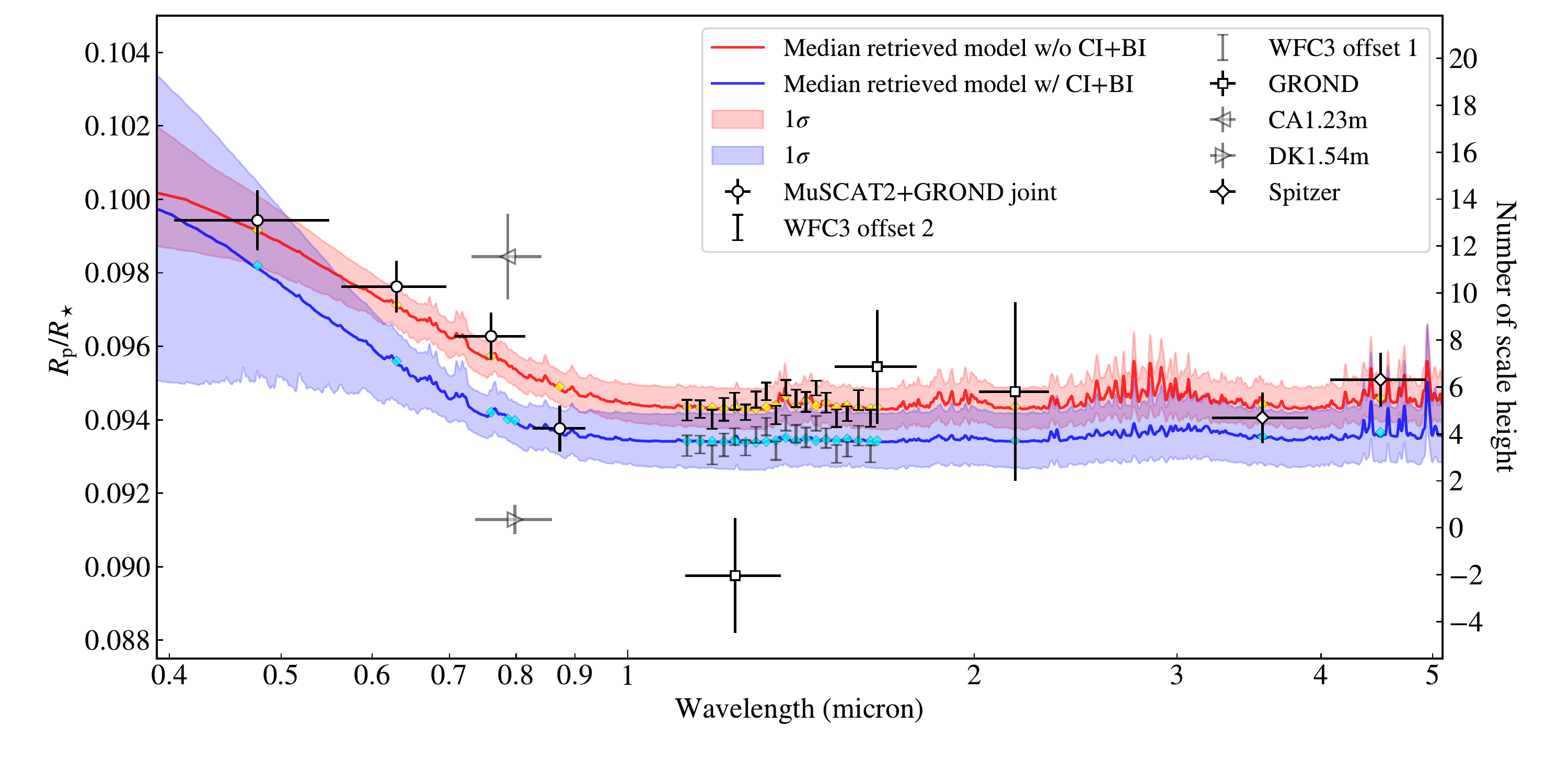}
\caption{Radius ratios in different passbands, including MuSCAT2 and GROND joint measurements (circles), GROND near-infrared observations \citep[squares;][]{Mancini2019MNRAS.485.5168M}, Calar Alto 1.23~m and Danish 1.54~m \citep[triangles;][]{Mancini2019MNRAS.485.5168M}, HST/WFC3 observations (no symbol) from \citet{Tsiaras2018AJ....155..156T}, and {\it Spitzer} measurements (diamonds). Blue line shows the median retrieved atmospheric model based on all measurements and the light blue band its $1\sigma$ uncertainty. Red line and its shaded band show the retrieval without the measurements in the Cousins-I and Bessell-I bands. The small diamonds in cyan and yellow colours show the binned version of the retrieved model in the passbands with observations. HST/WFC3 measurements were shifted using the best fit values from the two retrieval runs.} \label{fig:atmosphere}
\end{figure*}

Multi-colour observations of hot Jupiters can be used to construct a transmission spectrum, valuable to probe their atmospheres at the terminator. Figure~\ref{fig:atmosphere} shows the measured radius ratio in different passbands from the new MuSCAT2 observations together with our reanalysis of the observations from \textit{Spitzer}, \citet{Mancini2019MNRAS.485.5168M}, and the \citet{Tsiaras2018AJ....155..156T} {\it HST} observations. Our results disagree with those from \citet[][see their Figs.~8 and 9]{Mancini2019MNRAS.485.5168M}. We find no evidence of TiO/VO in the atmosphere of WASP-74~b, but a steep slope in the optical transmission spectrum. 

We attribute the origin of this disagreement to the different analysis of the light curves. First, we fit all the available light curves as a function of passband jointly, which constrains the geometry and limb darkening parameters better than modelling the light curves individually. Second, some of the GROND covariates are relatively noisy and contain a significant amount of strong outliers (especially the centroid estimates in the NIR passbands). Linear baseline models (the approach used also by \citealt{Mancini2019MNRAS.485.5168M}) do not perform well with noisy covariates, so we remove the photometry points where the covariates are clearly problematic (that is, we remove part of the data based on the covariates, but not on the photometry itself). This approach improves the baseline model significantly, and the separate MuSCAT2 and GROND analyses agree with each other well.

\begin{table}
\centering
{\renewcommand{\arraystretch}{1.2}
\footnotesize
\caption{Best fit parameters from the atmospheric retrieval. The prior label $\mathcal{U}$ represents a uniform distribution.}
\label{tab:atm_res}
\begin{tabular}{llcr}
\hline\hline
\noalign{\smallskip}
Parameter & Prior & Run 1 \tablefootmark{(a)} & Run 2 \tablefootmark{(b)}\\
\noalign{\smallskip}
\hline
\noalign{\smallskip}
$R_\mathrm{1bar}$ $[R_{\rm Jup}]$       & $\mathcal{U}(0.5R_\mathrm{p},1.5R_\mathrm{p})$    & $ 1.248^{+0.034}_{-0.036}$    & $ 1.274^{+0.018}_{-0.020}$\\
$T$ [K]                                 & $\mathcal{U}(0.5T_\mathrm{eq},1.5T_\mathrm{eq})$  & $ 1882^{+429}_{-510}$         & $ 1568^{+370}_{-306}$\\
Scatter slope $s$                       & $\mathcal{U}(-4,20)$                              & $ 14.8^{+3.5}_{-5.7}$         & $ 16.2^{+2.6}_{-4.2}$\\
$\log{f_{\rm scatter}}$                 & $\mathcal{U}(-10,10)$                             & $ 2.4^{+1.5}_{-2.4}$          & $ 3.0^{+1.3}_{-1.4}$\\
C/O ratio                               & $\mathcal{U}(0.05,2)$                             & $ 0.92^{+0.66}_{-0.56}$       & $ 0.60^{+0.53}_{-0.33}$\\
$\log{Z/Z_\odot}$                       & $\mathcal{U}(-1,3)$                               & $ 0.50^{+1.24}_{-0.98}$       & $ 0.37^{+1.07}_{-0.90}$ \\
$P_\mathrm{Cloud-top} [\log {\rm Pa}]$  & $\mathcal{U}(-0.99,5)$                            & $ 0.9^{+1.4}_{-1.2}$          & $ 1.3^{+1.1}_{-1.2}$ \\
WFC3 offset  [ppm]                      & $\mathcal{U}(-2000,2000)$                         & $434^{+131}_{-139}$           & $ 615^{+98}_{-97}$ \\
Error multiple                          & $\mathcal{U}(0.1,10)$                             & $ 2.3^{+0.4}_{-0.3}$          & $ 1.3^{+0.2}_{-0.2}$ \\
\noalign{\smallskip}
\hline
\end{tabular}}
\tablefoot{
  \tablefoottext{a}{Retrieval run on all available measurements.}
  \tablefoottext{b}{Retrieval run on measurements excluding Cousins-I and Bessell-I passbands.}
}
\end{table}

We used the PLanetary Atsmopheric Transmission for Observer Noobs code \texttt{PLATON}\footnote{\url{https://github.com/ideasrule/platon}} \citep{PLATON} to retrieve the atmospheric properties of WASP-74~b. \texttt{PLATON} is a fast, user-friendly open-source code for retrieval and forward modelling of exoplanet atmospheres written in Python. For our retrieval analysis, we assumed an isothermal atmosphere and did not take into account any contamination from stellar heterogeneities \citep[e.g.,][]{2014A&A...568A..99O,2018ApJ...853..122R,2019AJ....157...96R}. We fit the combined transmission spectrum, including the low-resolution HST/WFC3 spectrum from  \citet{Tsiaras2018AJ....155..156T}, reanalysed Cousins $I$, Bessell $I$, and GROND measurements from \citet{Mancini2019MNRAS.485.5168M}, and our new MuSCAT2 and {\it Spitzer} measurements. The free parameters are isothermal temperature $T$, planet's radius at a pressure of 1~bar $R_\mathrm{1bar}$, C/O ratio, atmospheric metallicity relative to the solar value $\log{Z/Z_\odot}$, scattering slope $s$, scattering amplitude $\log f_\mathrm{scatter}$, cloud-top pressure $\log P_\mathrm{cloud}$, and an error multiple to scale measurement uncertainties. We also allowed the WFC3 spectrum to have an overall free offset, as it was derived with a different set of $i$ and $a/R_\star$, which could introduce an offset due to the correlation between $R_\mathrm{p}/R_\star$ and those two transit parameters.

We performed two runs of retrieval analyses: one including all available measurements, while the other excluding Cousins $I$ and Bessell $I$ passbands. Figure~\ref{fig:atmosphere} shows the combined transmission spectrum along with the best retrieved 1$\sigma$ confidence region. In both retrieval runs, the retrieved transmission spectrum is almost featureless, except for the significant scattering slope in the optical. The retrieved atmospheric parameters are given in Table~\ref{tab:atm_res}, most of which are not well constrained by the current observations. The reported errors do not account for the errors in the input parameters, but only for the fitting procedure. The analysis favours a low value for the cloud-top pressure, which is consistent with the lack of water absorption feature in the \textit{HST}/WFC3 band. The analysis tends to retrieve an unconstrained scattering slope that always skews to the upper boundary. If the optical measurements are directly fitted by a linear function, the observed slope $s_\mathrm{obs}=-\mathrm{d}(R_\mathrm{p}/R_\star)/\mathrm{d}(\ln\lambda)$ can be converted to a scattering slope of $s=s_\mathrm{obs}R_\star/(k_\mathrm{B}T_\mathrm{eq}/\mu/g_\mathrm{p})=20.0\pm9.5$ or $14.1\pm 2.8$ for cases with or without Cousins $I$ and Bessell $I$, respectively, using $R_\star$ from Table~\ref{tab:star}, $T_\mathrm{eq}$ from Table~\ref{tab:parameters}, $g_\mathrm{p}$ from \citet{Mancini2019MNRAS.485.5168M} and assuming a mean molecular weight of $\mu=2.3$~g\,mol$^{-1}$. However, since different temperatures are retrieved from the two runs, the retrieved scattering slopes are correspondingly deviating from the estimates that assume the equilibrium temperature. The inclusion of Cousins $I$ and Bessell $I$ also degrade the goodness of fitting, which is dictated by the very small error bar of Bessell $I$.

The retrieved atmospheric models might be a challenge for theories. The observed "super-Rayleigh" slope ($s > 4$) could be explained by photochemical haze particles produced in a vigorously mixing atmosphere, where a steep positive opacity gradient relative to altitude can be achieved \citep{Kawashima2019ApJ...877..109K,Ohno2020ApJ...895L..47O}. However, the atmospheric temperature of this planet ($T \sim 1900$K) can be too high to sustain the hydrocarbon hazes. On the other hand, some mineral condensates that can exist in the hot atmosphere, such as Mg$_2$SiO$_4$, have refractive properties that can produce super-Rayleigh slopes without such opacity gradient \citep[e.g.,]{Wakeford2015A&A...573A.122W}. However, it may require a somewhat extreme condition (e.g., small particle size, namely high nucleation rate, and high atmospheric diffusivity) to reproduce the steepness of the slope as shown for the case of MgSiO$_3$ clouds in \citet{Ormel2019A&A...622A.121O}. 

 In addition, at the same time with the super-Rayleigh slope, it is also necessary to explain the flat spectrum in the near-infrared region observed by HST/WFC3. It is difficult to reproduce these two features only with a single-aerosol layer, and two (or more) aerosol layers are probably required, where the lower one has a thick grey opacity to reproduce the NIR flat spectrum and the upper one is composed of diffused aerosols that responsible for the super-Rayleigh slope \citep{Ehrenreich2014A&A...570A..89E,Dragomir2015ApJ...814..102D,Sing2015MNRAS.446.2428S}. This idea however requires very different diffusivities for the two layers, and it is uncertain whether such a condition is realistic or not.

 In any case, the current observations are not adequate for further detailed discussions, and additional observations with a higher precision, wider wavelength coverage, and/or higher spectral resolution would be essential to confirm and further characterise the enigmatic spectral features observed in this study.

\subsection{High-resolution transmission spectroscopy} \label{subsec:transmission}

\begin{figure}
\centering
\includegraphics[width=0.98\hsize]{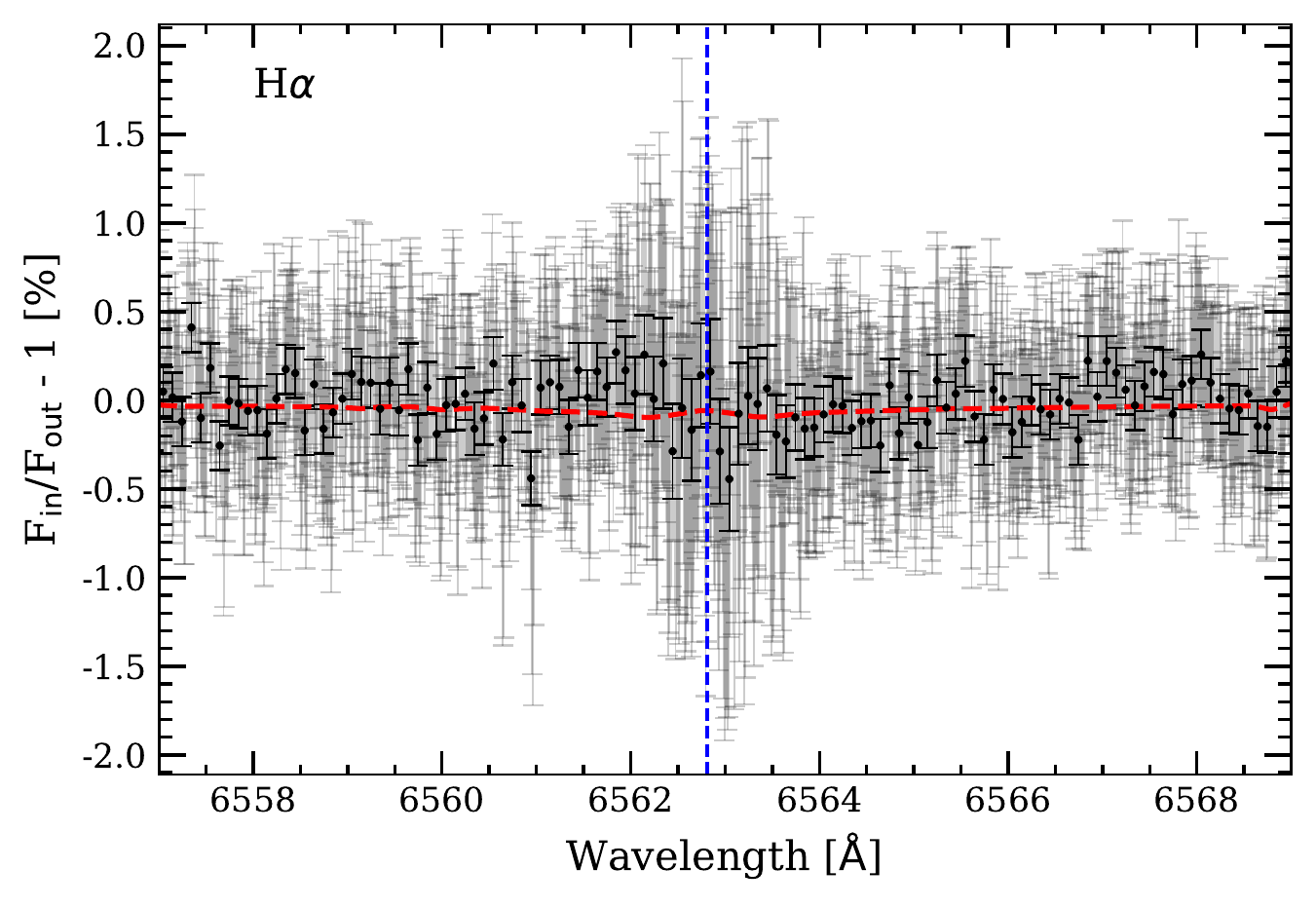}
\caption{Transmission spectrum of WASP-74~b in the H$\alpha$ line, combining the three nights observed with the HARPS-N spectrograph. In light grey we show the original result. The black dots correspond to the data binned by $0.1~{\rm \AA}$. The error bars come from the propagated photon noise.} \label{fig:TS_HR_Ha}
\end{figure}

\begin{figure*}
\centering
\includegraphics[width=0.98\hsize]{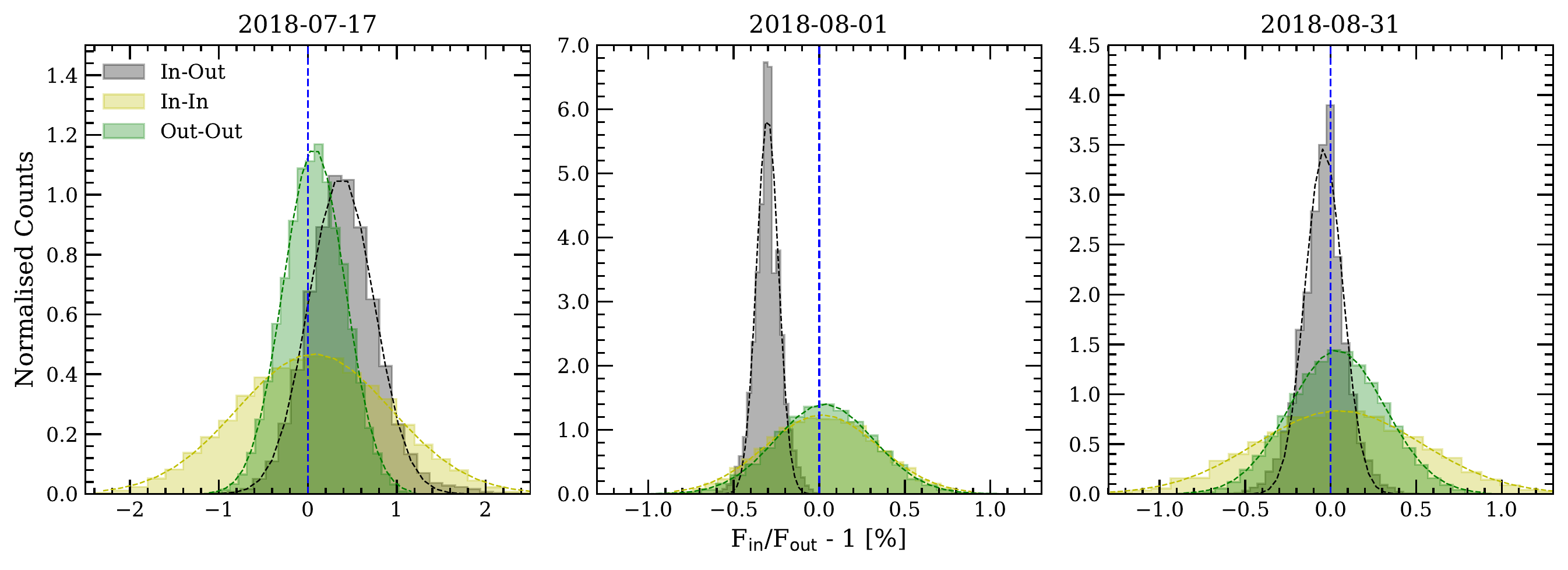}
\caption{Distributions of the EMC analysis in the H$\alpha$ line, using $20~000$ iterations and measuring the absorption depth with a bandwidth of $0.75~{\rm \AA}$. Each panel corresponds to the analysis of one night. In green we present the 'out-out' scenario, in yellow the 'in-in', and in grey the 'in-out', which corresponds to the atmospheric absorption scenario. The blue dashed vertical line marks the zero absorption level.} \label{fig:EMC_Ha}
\end{figure*}

High-resolution transmission spectroscopy observations are an excellent tool to study the atmospheric composition of exoplanets orbiting bright host stars \citep{Wytten2015A, Wytt2017A&A...602A..36W, Seidel2019arXiv190200001S, Cauley2019}. 

The one-dimensional HARPS-N spectra were corrected of the Earth telluric absorption contamination using Molecfit \citep{Molecfit1,Molecfit2}, as described in \citet{Allart2017} and used in recent atmospheric studies such as \citet{Hoeijmakers2018,Casasayas2019,Hoeijmakers2019}. After this correction, we follow the standard method to extract the atmospheric transmission spectrum \citep[see e.g.,][for details]{Wytten2015A,Casasayas2018,YanKELT9,Yan2019}. 

In particular, in order to move the spectra to the stellar rest frame, we use the stellar velocity semi amplitude ($K_{\star} = 114.1$\,m\,s$^{-1}$) measured by \citet{Hellier2015AJ....150...18H}. The master out-of-transit spectrum is computed by combining the out-of-transit data using the S/N of the order as weight. After computing the ratio of the spectra by this master out-of-transit spectrum, the residuals are moved to the planet rest frame using a planet velocity semi amplitude $K_p=178.92$\,km\,s$^{-1}$, derived from $K_{\star}$, and the planetary and stellar masses measured in this work (see Table~\ref{tab:star}). Finally, due to the long ingress and egress duration of the transit, only around 4 spectra were taken between the second and third contacts. For this reason, when computing the transmission spectrum, we average the spectra between the first and fourth contacts of the transit. We note that the selection of the in- and out-of-transit observations is performed using the transit epoch measured in Section~\ref{sec:rossitter}. 

We apply this method to different lines of the spectrum. In the case of \ion{Na}{i}, the S/N in the stellar lines core is too low to retrieve any atmospheric signature. For the first night, for example, the central core is at null counts level. Focusing on H$\alpha$ ($6562.81\,{\rm \AA}$; \citealt{NIST_ASD}), we combine the results of the three transit observations by using the mean S/N of each night in H$\alpha$ order as weights. The transmission spectrum is presented in Fig.~\ref{fig:TS_HR_Ha}. The results from the individual nights are presented in the Appendix for completeness (Fig.~\ref{fig:TS_ind}).

We also model the centre-to-limb variation (CLV) and RM effects in order to estimate the impact of both effects. For the CLV estimation, we follow \cite{Yan2017A&A...603A..73Y}, and also include the RM effect on the stellar lines profile as presented in \citet{YanKELT9} and \citet{Casasayas2019}. For this computation, the stellar spectra are modelled using VALD3 line list \citep{VALD3}, and MARCS \citep{MARCS2008A&A...486..951G} models, assuming solar abundance, local thermodynamic equilibrium and the stellar parameters presented in Section~\ref{sec:stellarparams}. For the WASP-74 system these effects have an impact smaller than ${\sim 0.1\%}$ (in relative flux) in the H$\alpha$ line core, which is included in error bars of the resulting transmission spectrum (see Fig.~\ref{fig:TS_HR_Ha}).  

Thus, we are not able to detect any feature with atmospheric origin. The achieved S/N of the spectra is very low at the line cores, specially for deeper lines such as \ion{Na}{i}. For this reason, and due to the equilibrium temperature of the planet ($1860~{\rm K}$), which is close to the ultra-hot Jupiter zone, our study is focused on the H$\alpha$ line. The transmission spectrum around this spectral line does not show any clear signature from the exoplanet atmosphere. In order to account for possible systematic effects, we compute the Empirical Monte Carlo analysis described in \citet{2008Redfield}. Figure~\ref{fig:EMC_Ha} shows that the "in-in" and "out-out" distributions are centred at zero absorption depth for all nights, as expected. On the other hand, the "in-out" distribution, which corresponds to the absorption scenario, is centred at a different position depending on the night. For the first and last nights, the absorption scenario can not be disentangled from the noise level due to the S/N achieved during the observations. However, for the second night where the S/N is the highest, the "in-out" distribution is centred at ${\sim}-0.3\%$. We note that with a transit duration of $2.38\,{\rm h}$ and using $600$\,s of integration per exposure, we are only able to measure around five spectra fully in-transit with relatively low S/N. The magnitude of the host star and its transit duration make WASP-74~b a challenging planet for atmospheric studies using $3.5$\,m telescopes. However, it is an ideal target to be studied using high-resolution spectrographs located on larger telescopes, such as ESPRESSO at the Very Large Telescope in Chile.

\section{Summary} \label{sec:discussion}

The obliquity of the WASP-74 system is measured for the first time, using three transits observed with the HARPS-N spectrograph. Here, we measure an aligned system with a projected spin-orbit angle of $0.8\pm1.0$ degrees, in agreement with previous findings suggesting that planets orbiting stars with effective temperatures lower than ${\sim 6200}~{\rm K}$ are aligned.

We further have used multi-colour observations of WASP-74~b to construct a transmission spectrum in order to probe its atmosphere. Our results disagree with those from \citet{Mancini2019MNRAS.485.5168M}, as we find no evidence of higher absorption in the bluer wavelengths, but a steep slope in the optical transmission spectrum. The origin of this disagreement is attributed to the different analysis of the light curves. We used the \texttt{PLATON} code to retrieve the atmospheric properties of WASP-74~b. We fit the combined transmission spectrum, including the low-resolution HST/WFC3 spectrum from  \citet{Tsiaras2018AJ....155..156T}, the GROND measurements and our new MuSCAT2 data. The retrieved transmission spectrum is almost featureless, except for the significant scattering slope in the optical.

Finally, using three transit observations of WASP-74~b with the HARPS-N spectrograph, we investigate its high-resolution transmission spectrum. Unfortunately, due to the low S/N of the data, we are not able to detect any feature with atmospheric origin. The magnitude of the host star and its transit duration make WASP-74~b a challenging planet for atmospheric studies using $4$-m class telescopes, but it is an interesting target to be further studied using high-resolution spectrographs placed in larger aperture telescopes.

\begin{acknowledgements}

This article is partly based on observations made with the MuSCAT2 instrument, developed by ABC, at Telescopio Carlos Sánchez operated on the island of Tenerife by the IAC in the Spanish Observatorio del Teide. It is also based on observations made with the Italian Telescopio Nazionale Galileo (TNG) operated on the island of La Palma by the Fundación Galileo Galilei of the INAF (Istituto Nazionale di Astrofisica) at the Spanish Observatorio del Roque de los Muchachos of the Instituto de Astrofisica de Canarias. 

R.\,L. has received funding from the European Union’s Horizon 2020 research and innovation program under the Marie Skłodowska-Curie grant agreement No.~713673 and financial support through the “la Caixa” INPhINIT Fellowship Grant LCF/BQ/IN17/11620033 for Doctoral studies at Spanish Research Centres of Excellence from “la Caixa” Banking Foundation, Barcelona, Spain. G.\,C. acknowledges the support by the B-type Strategic Priority Program of the Chinese Academy of Sciences (Grant No. XDB41000000) and the Natural Science Foundation of Jiangsu Province (Grant No. BK20190110). This work is partly financed by the Spanish Ministry of Economics and Competitiveness through grants ESP2013-48391-C4-2-R. 
This work is supported by JSPS KAKENHI grant Nos. 18H05442, 15H02063, 22000005, JP17H04574, JP18H01265, and JP18H05439, and JST PRESTO Grant Number JPMJPR1775.

\end{acknowledgements}

\bibliographystyle{aa} 
\bibliography{biblio} 

\onecolumn
\begin{appendix} 

\section{High-resolution transmission spectroscopy. Additional results}

\begin{figure}[h]
\centering
\includegraphics[width=0.49\hsize]{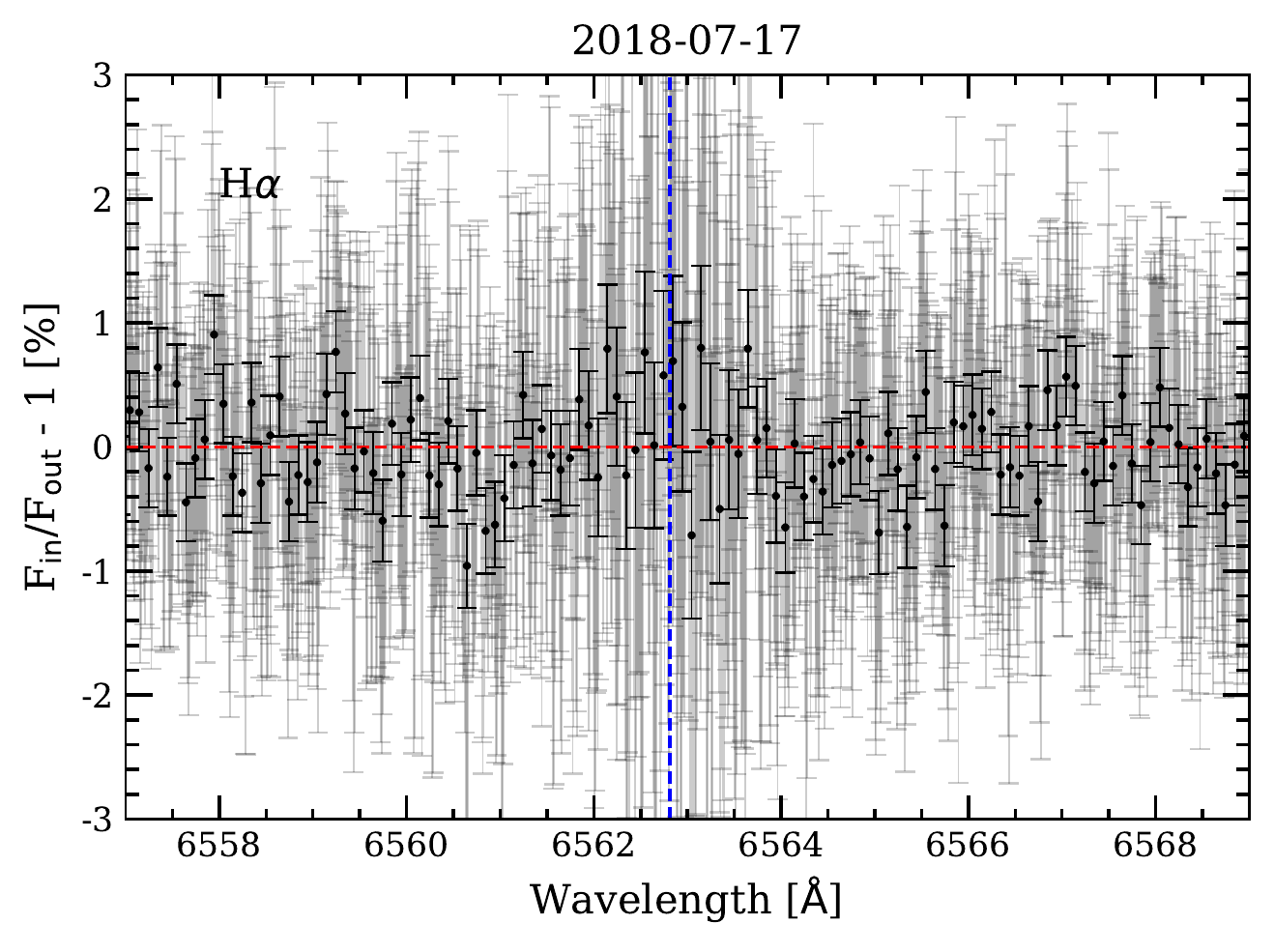}
\includegraphics[width=0.49\hsize]{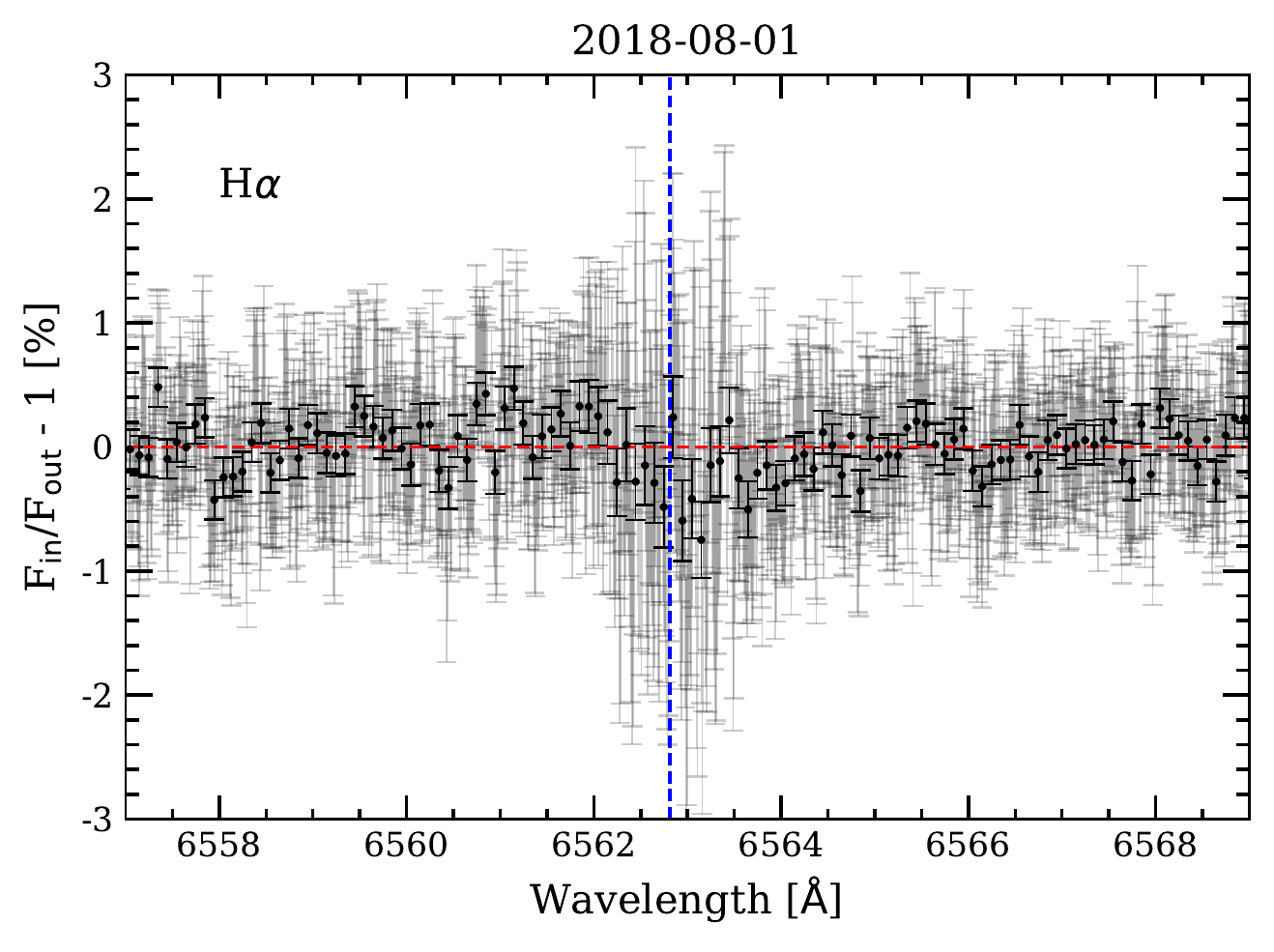}
\includegraphics[width=0.49\hsize]{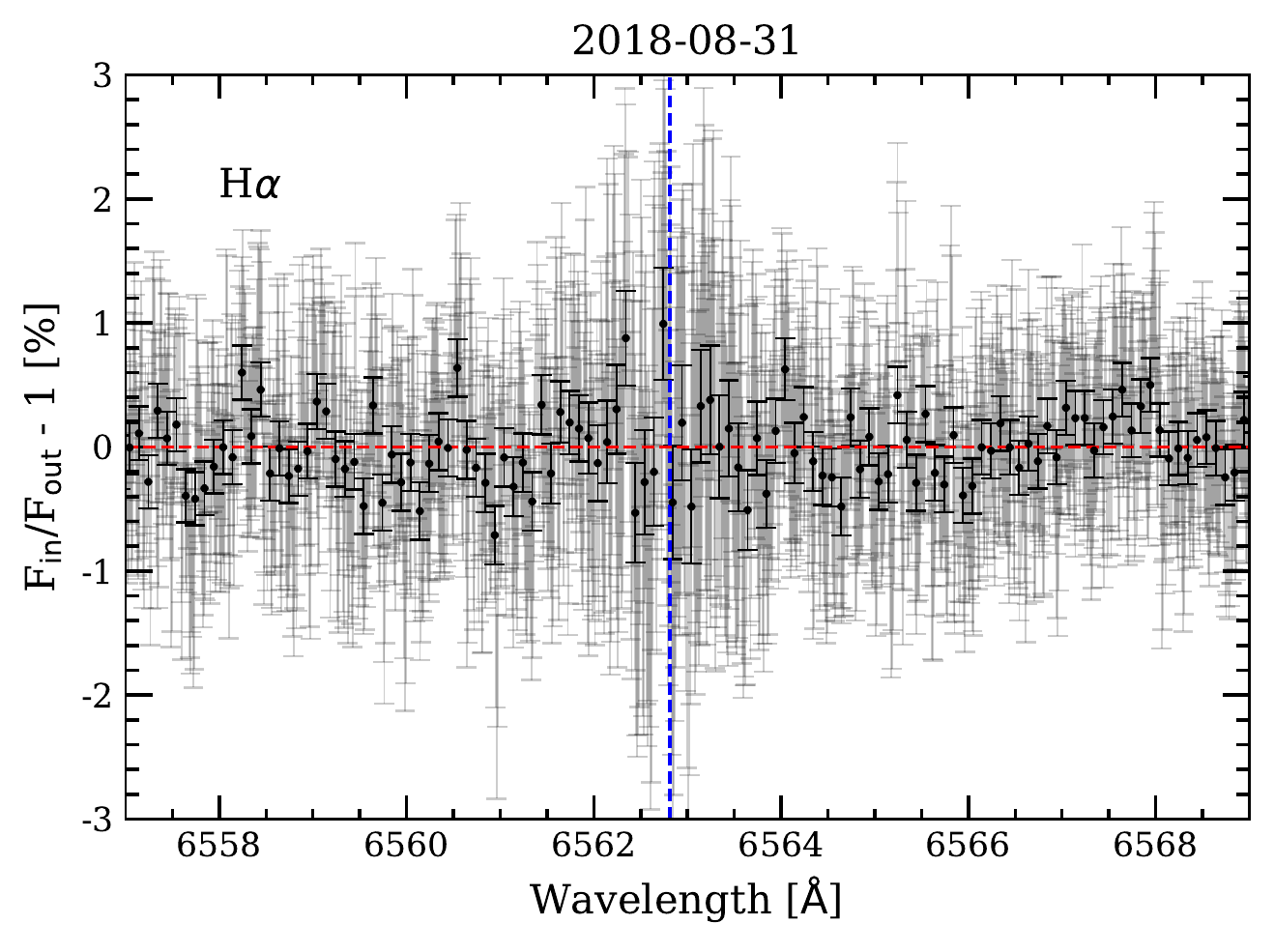}
\caption{Individual transmission spectra around H$\alpha$ line. In light gray we show the original data, while in black dots the data is binned by $0.1~{\rm \AA}$. In blue we show the H$\alpha$ laboratory position and in red the null absorption level.} \label{fig:TS_ind}
\end{figure}

\end{appendix}

\end{document}